\documentclass[iicol]{sn-jnl}

\usepackage{graphicx}
\usepackage{hyperref}
\hypersetup{hidelinks}
\usepackage{natbib}

\begin{document}

\title[Comparing nasal anatomies]{On the comparison between pre- and post-surgery nasal anatomies via computational fluid dynamics}


\author[1]{\fnm{Eric} \sur{Segalerba}}\email{eric.segalerba@edu.unige.it}
\author[2]{\fnm{Gabriele} \sur{Dini Ciacci}}\email{gabriele.diniciacci@polimi.it}
\author*[2]{\fnm{Maurizio} \sur{Quadrio}}\email{maurizio.quadrio@polimi.it}
\author*[1]{\fnm{Jan O.} \sur{Pralits}}\email{jan.pralits@unige.it}

\affil*[1]{\orgdiv{Department of Civil, Chemical and Environmental Engineering}, \orgname{University of Genova}, \orgaddress{\street{Via Montallegro, 1}, \city{Genoa}, \postcode{16145}, \country{Italy}}}

\affil*[2]{\orgdiv{Department of Aerospace Sciences and Technologies}, \orgname{Politecnico di Milano}, \orgaddress{\street{Campus Bovisa}, \city{Milano}, \postcode{20156}, \country{Italy}}}

\abstract{
Nasal breathing difficulties (NBD) are widespread and difficult to diagnose; the failure rate of their surgical corrections is high. 
Computational Fluid Dynamics (CFD) enables diagnosis of NBD and surgery planning, by comparing a pre-operative (pre-op) situation with the outcome of virtual surgery (post-op). 
An equivalent comparison is involved when considering distinct anatomies in the search for the functionally normal nose.
Currently, this comparison is carried out in more than one way, under the implicit assumption that results are unchanged, which reflects our limited understanding of the driver of the respiratory function.

The study describes how to set up a meaningful comparison.
A pre-op anatomy, derived via segmentation from a CT scan, is compared with a post-op anatomy obtained via virtual surgery. 
State-of-the-art numerical simulations for a steady inspiration carry out the comparison under three types of global constraints, derived from the field of turbulent flow control: a constant pressure drop (CPG) between external ambient and throat, a constant flow rate (CFR) through the airways and a constant power input (CPI) from the lungs can be enforced.
A significant difference in the quantities of interest is observed depending on the type of comparison. 
Global quantities (flow rate, pressure drop, nasal resistance) as well as local ones are affected.
The type of flow forcing affects the outcome of the comparison between pre-op and post-op anatomies. 
Among the three available options, we argue that CPG is the least adequate. Arguments favouring either CFR or CPI are presented.}


\maketitle

\section{Introduction}\label{sec:introduction}
Nasal breathing difficulties (NBD) are a widespread condition; it is well known \citep{gray-1978} that the large majority of population exhibits some anatomic deformity of the nasal airways.
Clinicians are frequently confronted with the issue of whether such deformities are the actual cause of the patient's symptoms: while some situations, e.g. an overly deviated septum, are self-evident, the interpretation of less pronounced deformities is often debatable.
The number of surgeries, and in general the burden induced by NBD on the healthcare system, is large.
Surgeons rely on their own judgment and experience to take surgical decisions, but errors are unavoidable \citep{rhee-etal-2014}.
The subjectivity of such choices leads to several unnecessary surgeries being performed each year worldwide: a large failure rate of the interventions actually carried out is recorded e.g. for septoplasty \citep{sundh-sunnergren-2015} or maxillectomy \citep{bertazzoni-etal-2017}. 

Numerical analysis, as a tool to investigate bio-mechanical problems, is becoming common practice in several areas, including the nasal airways, where computational fluid dynamics (CFD) is increasingly used in several studies: see e.g. \cite{inthavong-etal-2019} for a recent and authoritative review. 
CFD makes virtual surgery possible \citep{radulesco-etal-2020, moghaddam-etal-2020}, by enabling the comparison between the flow in the original (pre-surgery or pre-op) anatomy and the flow in the (post-surgery or post-op) anatomy modified by the surgeon on the computer.
However, even though computing the flow in the nasal cavities via CFD may seem straightforward, a clearly defined and standardized procedure is still lacking. Fundamental questions regarding the flow in the human nose remain unaddressed, reflecting our limited understanding of its physiological driver(s), and this has hindered so far a widespread use of CFD for clinical purposes. 

This contribution addresses one such question, which has not been identified so far, let alone answered: How should a comparison between pre-op and post-op anatomies be carried out? The question is apparently simple, yet the answer is non-trivial, and requires putting together concepts ranging from numerical analysis to physiology of the entire respiratory system. Once the scope of the question is enlarged to include the comparison of two generic anatomies, it becomes apparent that an appropriate answer is crucial for the successful identification of the functionally normal nose.

By surveying the existing literature, and limiting the analysis to the frequent case of steady inspiration (or expiration), one notices that several CFD simulations of the nose flow enforce either a constant pressure difference between the external ambient and some point in the trachea \citep[see e.g. ][]{cannon-etal-2013, radulesco-etal-2019, cherobin-etal-2020} 
or a certain flow rate through the passageways \citep[see e.g.][]{lindemann-etal-2013, calmet-etal-dmd-2020, bruening-etal-2020}).
The first choice does not appear to possess a clear physiological rationale, whereas the second implies a comparison under the constraint of the same oxygen consumption rate. 
Statistically, about $2/3$ of the papers employ the latter approach. The two choices will be shown here to be not equivalent; hence, one has to decide beforehand which global quantity is kept constant across a comparison. Moreover, a third option will be introduced.

Although in a vastly different context, the very same question was identified and answered by \cite{frohnapfel-hasegawa-quadrio-2012} in the field of turbulence and flow control. In that case, the complex nasal anatomy reduces to a much simpler duct (a straight channel or a pipe, for example): still, either a pressure drop must be established across the inlet and the outlet sections, or a flow rate must be imposed, for the fluid to flow through the duct. 
Surgery can be interpreted as flow control via shape optimization; the pre- and post-op anatomies correspond to the flow without and with flow control.
In a duct flow, the comparison can be carried out by either enforcing a Constant Pressure Gradient (CPG) and measuring the flow rate, or enforcing a Constant Flow Rate (CFR) and measuring the pressure drop. 
A third option, named Constant Power Input (CPI) \citep{hasegawa-quadrio-frohnapfel-2014}, was also proposed as a further alternative, in which the quantity that remains constant across the comparison is the pumping power that enters the system, given by the product of the pressure drop and the flow rate.
In an indefinite plane channel flow, the choice between CPG, CFR and CPI has been found to imply minor differences in the statistical description of the same flow \citep{quadrio-frohnapfel-hasegawa-2016}, and a similar result is expected for the nasal flow. 
The point of concern, though, is that the choice of the forcing term becomes crucial when flow control is applied to reduce the skin-friction aerodynamic drag, and two {\em different} flows (albeit geometrically similar) need to be compared: in this case, the outcomes of CFR, CPI and CPG simulations differ significantly.

The objective of the present paper is thus to assess whether or not comparing pre-op and post-op anatomies is affected by the choice of one among the CFR, CPI and CPG strategies. 
We will delineate a simple CFD setup, where for example the clinically important temperature field is not computed, and consider one patient-specific pre-op anatomy with a corresponding post-op anatomy already available. It was created with virtual surgery, specifically with an endoscopic medial maxillectomy, and has been described by \cite{saibene-etal-2020}, where it was used as a reference surgical approach to develop alternative options which partially preserve the middle turbinate.

The structure of the paper is as follows: in Sec. \ref{sec:methods} the anatomies are described, and details about the CT scan and its reconstruction are provided; the computational model (equations, discretization and treatment of turbulence) is illustrated, and the three types of flow forcing are described in detail. 
In Sec. \ref{sec:results} both quantitative and qualitative results obtained for the two anatomies with the different forcings are described and compared, including the first-ever CPI simulations of the flow in the human nose. 
Sec. \ref{sec:discussion} critically discusses the results in terms of the significance of the various comparisons; lastly, Sec. \ref{sec:conclusions} summarizes the study and indicates the importance of a clinical consensus to define the best comparison strategy.

\section{Methods}
\label{sec:methods}

\subsection{Anatomy and computational domain}
The pre-op anatomy considered in the present study is obtained from segmentation of a CT scan. The post-op anatomy, instead, is built after a virtual maxillectomy of the former. Both have been described and discussed at length by \cite{saibene-etal-2020}. 

\begin{figure*}
\centering
\includegraphics[width=0.8\textwidth]{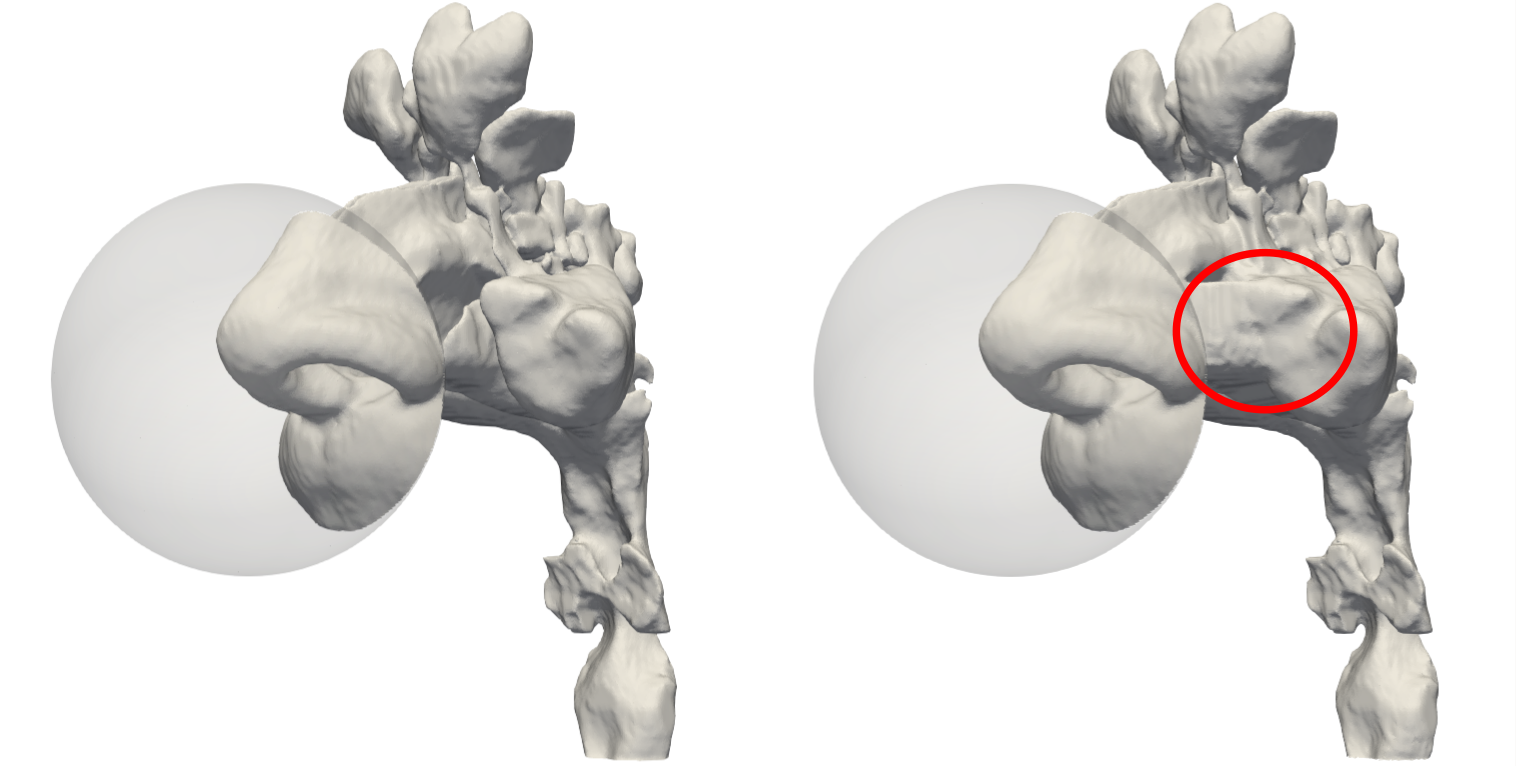}
\caption{Pre-op (left) and post-op (right) anatomies, including an external spherical volume arond the nose tip. In the post-op anatomy, the red circle highlights changes due to the virtual maxillectomy.}
\label{fig:STL}
\end{figure*}

The CT scan of a 67-year-old man, consisting of 384 DICOM images with spatial and coronal resolution of $0.5 \ mm$ and an axial gap of $0.6 \ mm$, is segmented, at constant radiodensity threshold, under supervision of an ENT expert, according to a previously described procedure \citep{quadrio-etal-2016}. 
Figure~\ref{fig:STL} portraits the pre-op anatomy, complemented by a spherical air volume surrounding the external nose, designed to move the inlet portion of the computational domain far from the nostrils while minimizing the computational overhead. 

The post-op anatomy is obtained by virtual surgery of the same patient for endoscopic medial maxillectomy, executed under close guidance of an ENT surgeon \citep{saibene-etal-2020}. 
This surgery is a standard procedure in the management of maxillary sinus neoplasms, and is sometimes employed to address inflammatory conditions. It has been chosen because it clearly alters the nasal resistance, and thus constitutes a convenient test bed for the present study.

\subsection{Computational procedures}
The pre- and post-op flow fields for a steady-state inspiration are computed with a relatively standard high-fidelity CFD approach, shortly described below. 
For the numerical solution of the flow equations, we employ the open-source finite-volumes solver OpenFOAM \citep{weller-etal-1998}. 
The computational domain shown in Figure~\ref{fig:STL} is discretized into a volume mesh which contains approximately 14.6 millions cells for the pre-op anatomy, and 15.4 millions for the post-op one, whose volume is slightly larger. 
Meshing starts from an uniform background mesh of cubic cells, with edge length of 250 microns, which is deformed and refined near the boundary in the process of adaptation to the curved boundary. The maximum non-orthogonality is less than $60^\circ$, and its mean value is $4.4^\circ$; maximum skewness is $3.2^\circ$.

In physiological conditions, the nasal flow is typically transitional with coexisting  laminar and subcritical turbulent regions. 
Moreover, it is often unsteady even when the boundary conditions are steady. 
Hence we adopt a time-resolved approach, i.e. a high-resolution Large-Eddy Simulation (LES), in which most of the turbulent flow scales are resolved, and only the smallest scales are modelled. 
The LES turbulence model is the Wall-Adapting Local Eddy-viscosity \citep{nicoud-ducros-1999}, which is able to turn itself off in regions where turbulence is absent. It should be realized, though, that with a high-resolution LES the turbulence model becomes relatively unimportant, since the mesh is fine enough to render the model contribution small or negligible.

The differential operators are all discretized at second-order accuracy, as it has been shown by \cite{schillaci-quadrio-2022} that a lower order deteriorates the solution to an unacceptable level, regardless of the turbulence modeling approach. 
The incompressible LES equations are solved with no-slip and no-penetration conditions applied to the solid boundaries; the boundary conditions enforced at the inflow (sphere) and at the outflow (throat) boundaries depend on the specific forcing, and will be discussed below in Sec. \ref{sec:forcing}. 
The temporal discretization uses a second-order implicit scheme: no stability constraint limits the size of the time step. However, for accuracy a value of the Courant–-Friedrichs–-Lewy number of $0.3$ is imposed in all cases. The average time step is $1.1 \cdot 10^{-5}$ seconds for the pre-op cases, and $4.5 \cdot 10^{-6}$ seconds for the post-op cases. 

Each simulation computes one second of physical time in about 4000 core hours. Parallel computing is used to reduce the computing time to less than two days.

\subsection{Flow forcing}
\label{sec:forcing}

The pre-op and post-op anatomies are compared for a case of steady inspiration, in which either the volumetric flow rate $Q$ (CFR), the pressure difference $\Delta p$ between inlet and outlet (CPG) or the power input $P$ entering the system (CPI) are kept constant across the comparison. 
(Note that, in an incompressible flow, pressure is defined within an arbitrary constant; a reference pressure $p=0$ is thus set at the outer ambient, and assigning $\Delta p$ becomes equivalent to assigning the pressure $p_{th}$ at the throat.)  
A close analogy exists between the present problem and the comparison of two turbulent duct flows, where flow control is used to alter the natural relation between the flow rate and the pressure drop, i.e. the friction law \citep{hasegawa-quadrio-frohnapfel-2014}. 

The first two options are simple from a practical point of view: in a typical CFD flow solver the numerical values for the quantities to be kept constant can be straightforwardly assigned. 
In CFR, one enforces a constant flow rate $Q=Q_0$, obtains a time-varying outlet pressure $p_{th}(t)$ and computes its average value $\overline{p}_{th}$ {\em a posteriori}.
In CPG, one enforces a constant outlet pressure $p_{th} = p_0$, obtains a time-varying flow rate $Q(t)$ and computes its average value $\overline{Q}$ {\em a posteriori}.

The third approach, CPI, is less conventional and typically not immediately available in standard CFD solvers, as power cannot usually be prescribed directly, but is at least as physically sound as the others. In CPI, one enforces a constant power input $P=P_0$, where $P$ is the product of the pressure drop and the flow rate:
\begin{equation}
P = Q \Delta p = - Q p_{th},
\label{eq:CPI}
\end{equation} 
in which the last identity follows from having set $p=0$ at the inlet. The solution then yields $p_{th}(t)$ and $Q(t)$, from which the values $\overline{p}_{th}$ and $\overline{Q}$ are computed both {\em a posteriori}.

In a time-dependent calculation, the simplest numerical implementation of Eq.(\ref{eq:CPI}) computes at each time instant $t$ the throat pressure $p_{th}^t$ needed to drive the flow as a function of the flow rate $Q^{t-dt}$ at the previous time $t - dt$, i.e.:
\begin{equation}
p_{th}^t = - \frac{P_0}{Q^{t-dt}} ,
\label{eq:CPIdisc}
\end{equation} 
where $dt$ is the time step. 

In analogy with the duct flow problem, it is important to notice that the three options are essentially equivalent as long as a single anatomy is considered.
In other words, a given case can be computed with CFR by enforcing a constant $Q=Q_0$ to obtain as a result a certain mean pressure $\overline{p}_{th}=p_0$ and mean power $\overline{P}=P_0$; the same case computed with CPG by enforcing a constant $p_{th}=p_0$ yields $\overline{Q}=Q_0$ and $\overline{P}=P_0$; and when CPI is used by enforcing a constant $P=P_0$, one gets $\overline{Q}=Q_0$ and $\overline{p}_{th}=p_0$. 
It is only when distinct anatomies are considered that differences may become significant. This is exactly the scenario considered in the present work.

\section{Results}
\label{sec:results}

\begin{figure}
\centering
\includegraphics[width=\columnwidth]{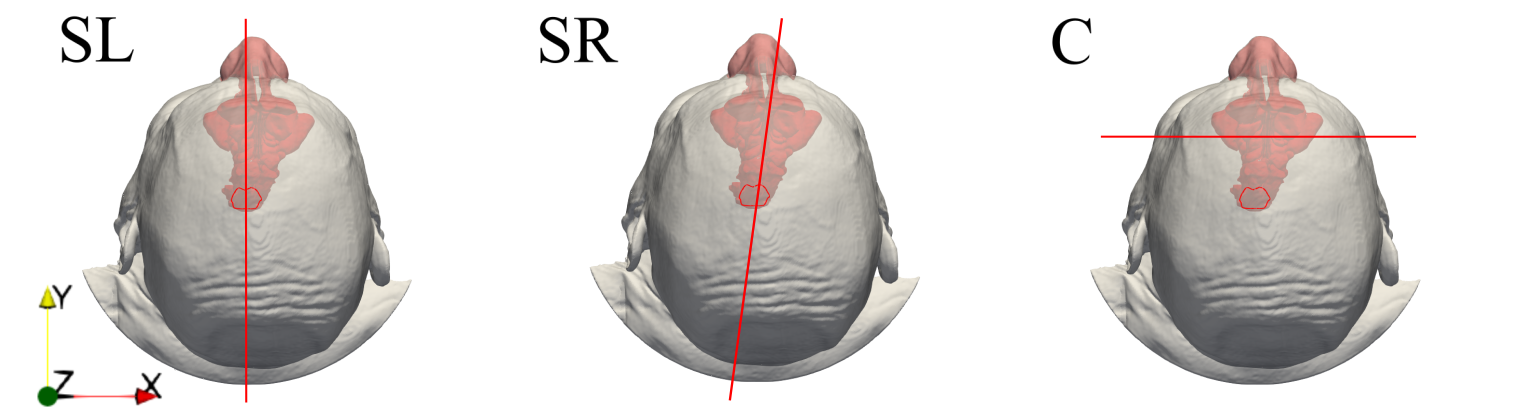}
\caption{Planes used throughout the paper to visualize results. Left: para-sagittal plane SL passing through the unaltered left nostril; center: para-sagittal oblique plane SR cutting through the operated right nostril and the throat; right: coronal plane C cutting through the maxillary sinuses. The $x$-axis is normal to the sagittal plane and points to the right; the $y$-axis is normal to the coronal plane and points towards the nose tip, and the $z$-axis is normal to the transverse plane and points upwards.}
\label{fig:planes}
\end{figure}

Results from six simulations are presented, comparing the pre-op and the post-op cases computed with the three options (CFR, CPG and CPI) available for the flow forcing. 
The simulations consider a steady inspiration with a flow rate of about $16 \ l/min$ or $2.67 \cdot 10^{-4} \ m^3/s$,  corresponding to a mild breathing intensity \citep{wang-lee-gordon-2012}. Equivalently, the reference case has $24.45 \ Pa$ of pressure difference between the external ambient and the throat, and $6.53 \ mW$ of mechanical power used to drive the flow through the nasal cavities.
The LES approach computes the temporal evolution of the flow; after excluding the initial transient, the time-dependent solution is averaged over one second of physical time to compute flow statistics. Based on previous experience \citep{covello-etal-2018}, we know that at these values of breathing rate averaging the time-dependent solution for 0.6 seconds is sufficient to obtain accurate flow statistics. This duration is almost doubled here, as the study aims at appreciating differences of flow statistics. 

Figure \ref{fig:planes} presents the three planes used in the following to illustrate and discuss the flow fields. We will consider two para-sagittal planes, the first (SL) cutting through the left, unaltered side of the airways, and the second (SR) cutting through the right side, modified by virtual surgery. Plane SR is not perpendicular to the $x$ axis, but is slightly inclined such that it passes through the throat. Lastly, a coronal plane (C) intersects the maxillary sinuses. 

\subsection{Constant Pressure Gradient (CPG)}
In a CPG simulation, the flow is driven by the pressure difference between the inlet (the external surface of the sphere, where the reference pressure is set to zero) and the outlet (the bottom plane in the throat region), directly enforced as a boundary condition as $p_{th}=p_0$. Assigning the throat pressure is a common practice in the CFD literature of the nasal airflow, see e.g. \cite{otto-etal-2017,farzal-etal-2019,li-etal-2019,plasek-etal-2022} among many others. 

The total flow rate increases from a pre-op value of $2.67 \cdot 10^{-4} \ m^3/s$ (or $16.02 \ l/min$) to a post-op value of $3.12 \cdot 10^{-4} \ m^3/s$ (or $18.72 \ l/min$) resulting in a percentage increase of 16.9\%. The corresponding increase in power input, defined by Eq.(\ref{eq:CPI}) is 16.9\%. 

\begin{figure}
\centering
\includegraphics[width=\columnwidth]{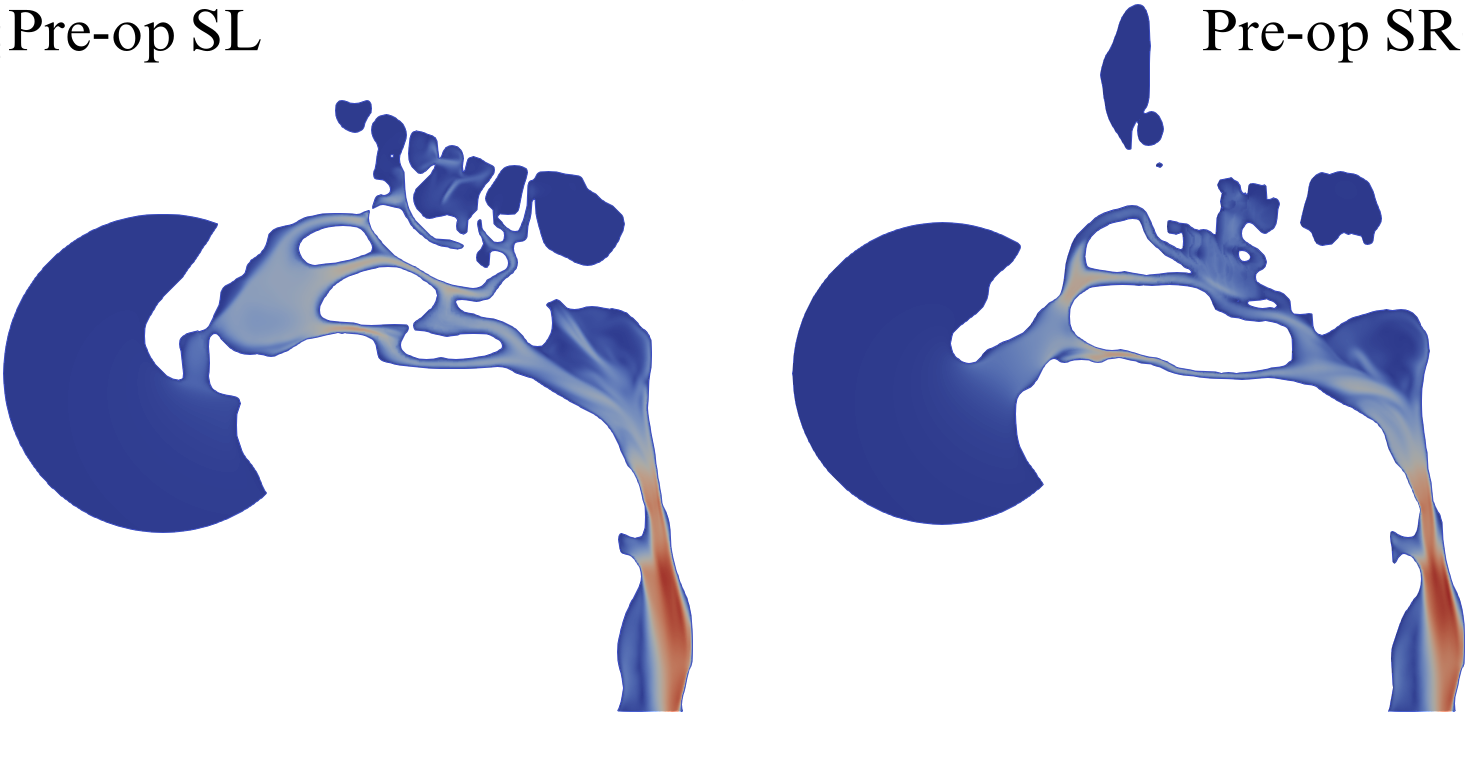}
\includegraphics[width=\columnwidth]{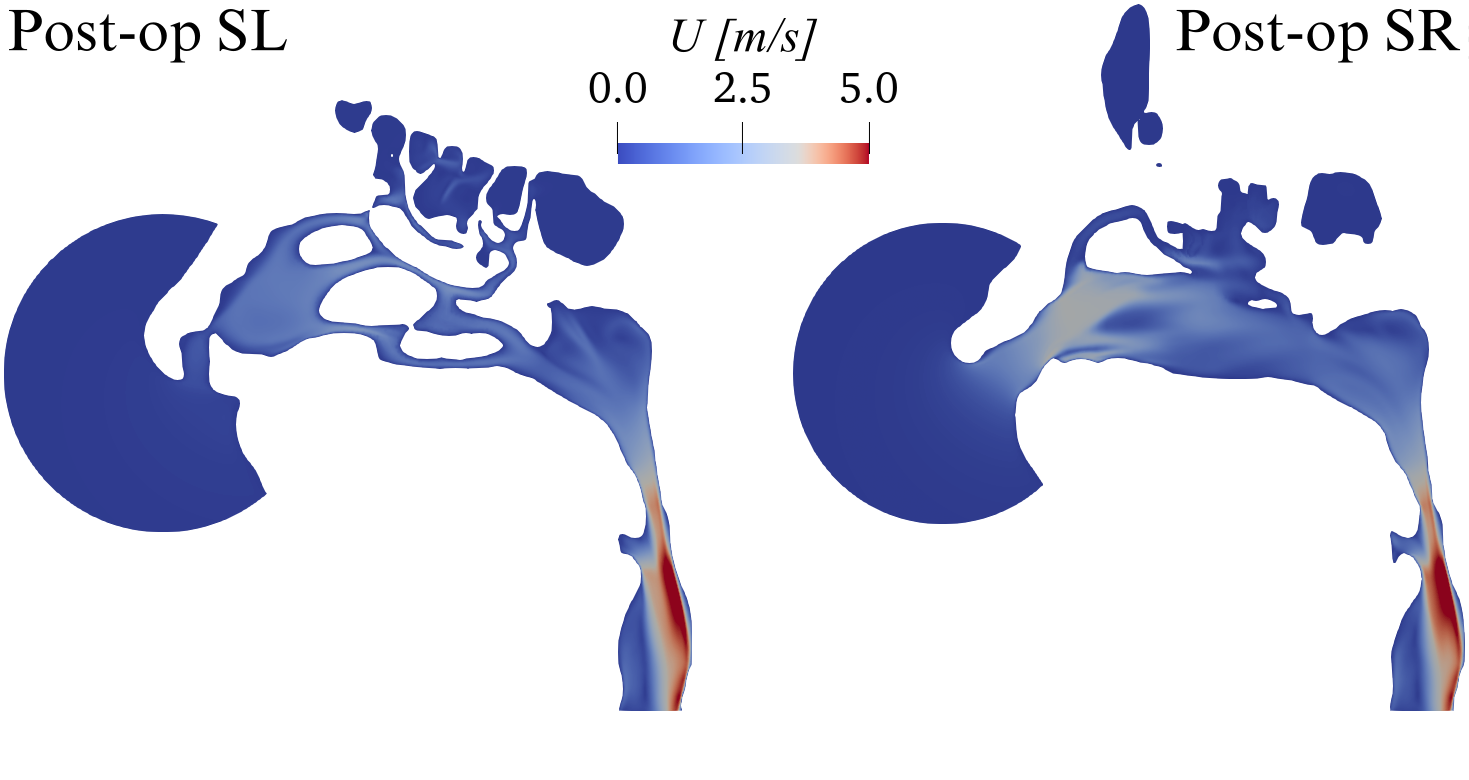}
\caption{CPG comparison: magnitude $U$ of the mean velocity vector in planes SL and SR.}
\label{fig:CPG-S}
\end{figure}

Figure~\ref{fig:CPG-S} compares the magnitude $U$ of the mean velocity for the pre-op and post-op anatomies in the para-sagittal planes SL and SR. Differences are minute in the unaltered SL side; the operated side, visible in the SR cut, presents large anatomical differences, as the maxillary sinus is removed; the flow field is obviously quite different as well.
Although qualitative differences can be discerned throughout the whole SR plane, the effect of the surgery is particularly evident in the vestibulum and at the nasopharynx, where the post-op flow shows a more uniform velocity distribution. 

\begin{figure}
\centering
\includegraphics[width=\columnwidth]{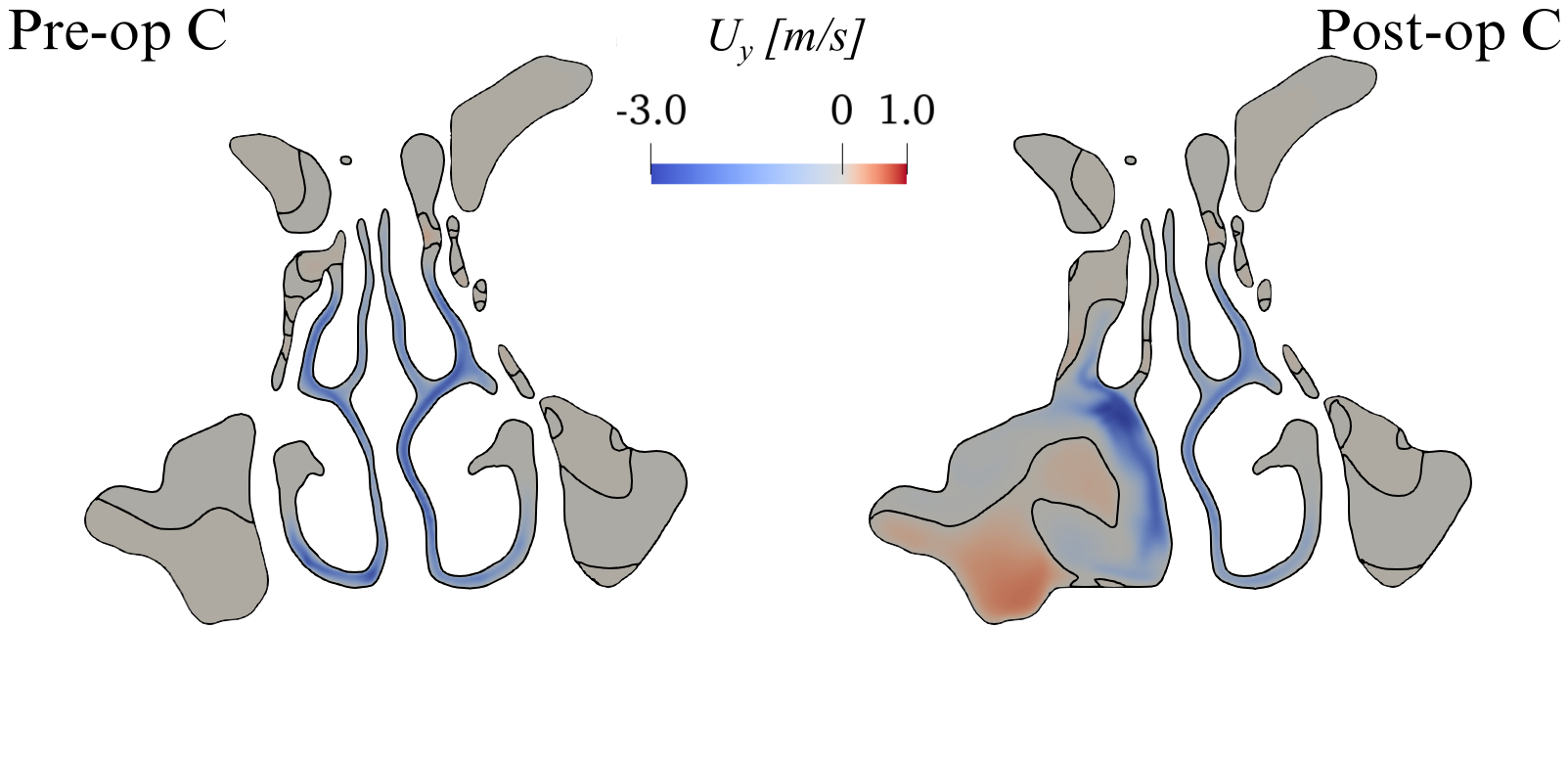}
\caption{CPG comparison: sagittal component $U_y$ of the mean velocity in the coronal plane C. The black line represents the zero contour level.}
\label{fig:CPG-C}
\end{figure}

An inspection of the mean sagittal velocity $U_y$ in the coronal plane, shown in  Figure~\ref{fig:CPG-C}, reveals that the virtual surgery has created an ample region of intense reverse flow, with air flowing backwards from the rhinopharynx towards the external ambient and reaching the remarkable reverse speed of $1 \ m/s$.

\subsection{Constant Flow Rate (CFR)}
In a CFR simulation, the flow is still driven by a pressure difference between inlet and outlet; however, its value is variable in time and adjusted during the simulation to achieve the target value of the flow rate $Q=Q_0$ which is enforced via the boundary condition. Assigning the flow rate is perhaps the most popular choice in the CFD literature of the nasal airflow, see e.g. \cite{lee-etal-2010,liL-etal-2019,vanstrien-etal-2021,berger-etal-2021}. 

The computed time-averaged pressure drop in the pre-op case is $24.45 \ Pa$, and as expected is identical to the one enforced in CPG. In the post-op case, the pressure drop reduces to $18.5 \ Pa$, with a decrease of 24.6\%. The corresponding decrease in power input is 24.6\%. 

\begin{figure}
\centering
\includegraphics[width=\columnwidth]{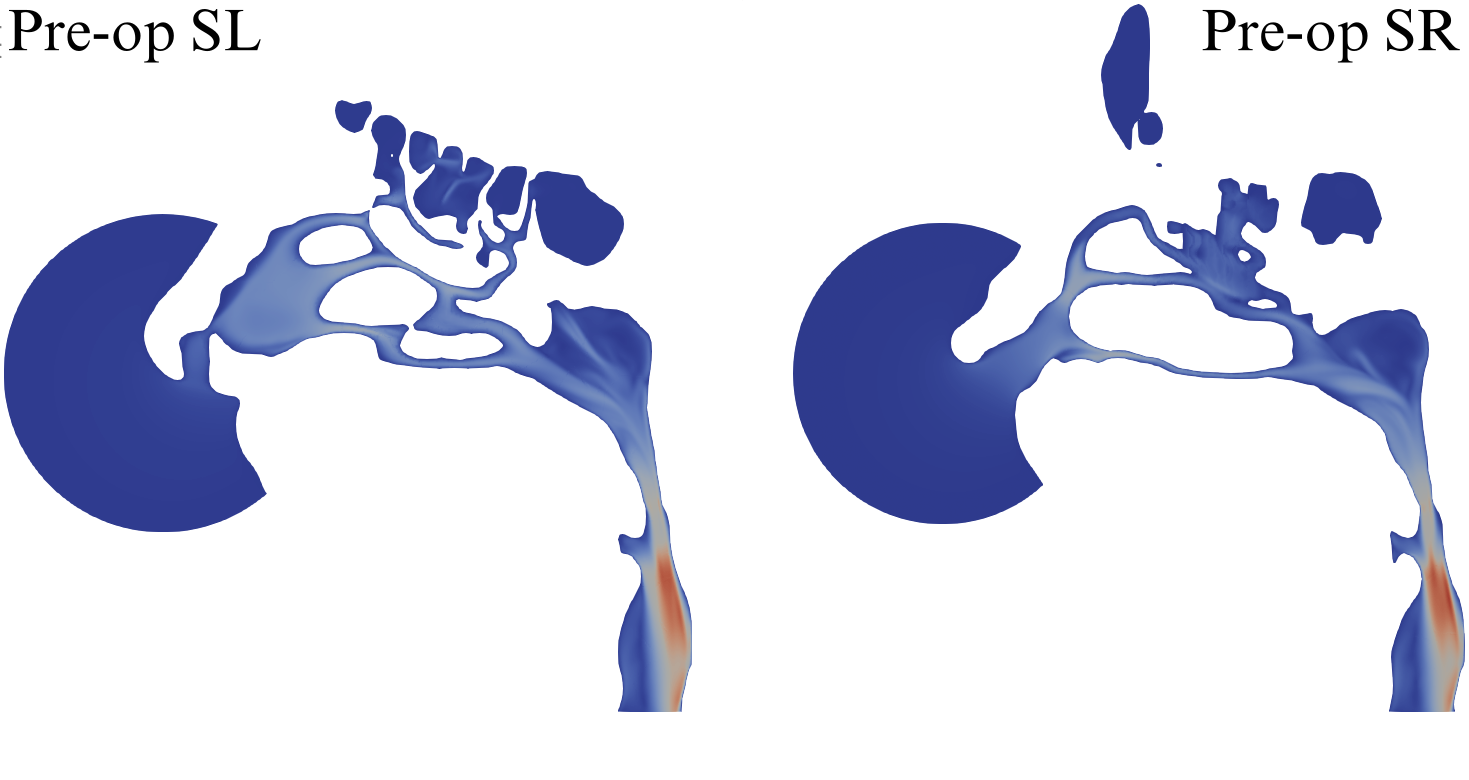}\\
\includegraphics[width=\columnwidth]{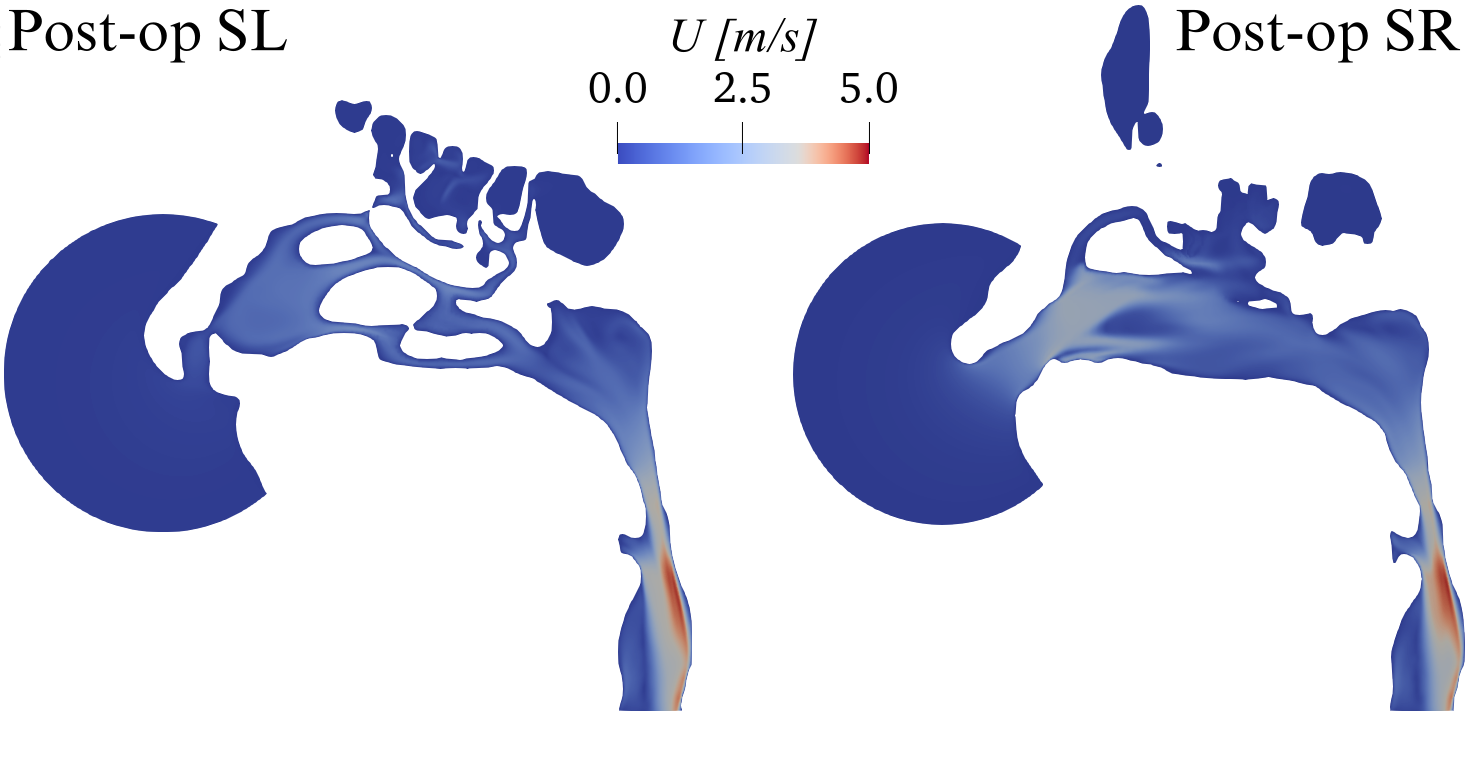}\\
\caption{CFR comparison: magnitude $U$ of the mean velocity vector in planes SL and SR.}
\label{fig:CFR-S}
\end{figure}

\begin{figure}
\centering
\includegraphics[width=\columnwidth]{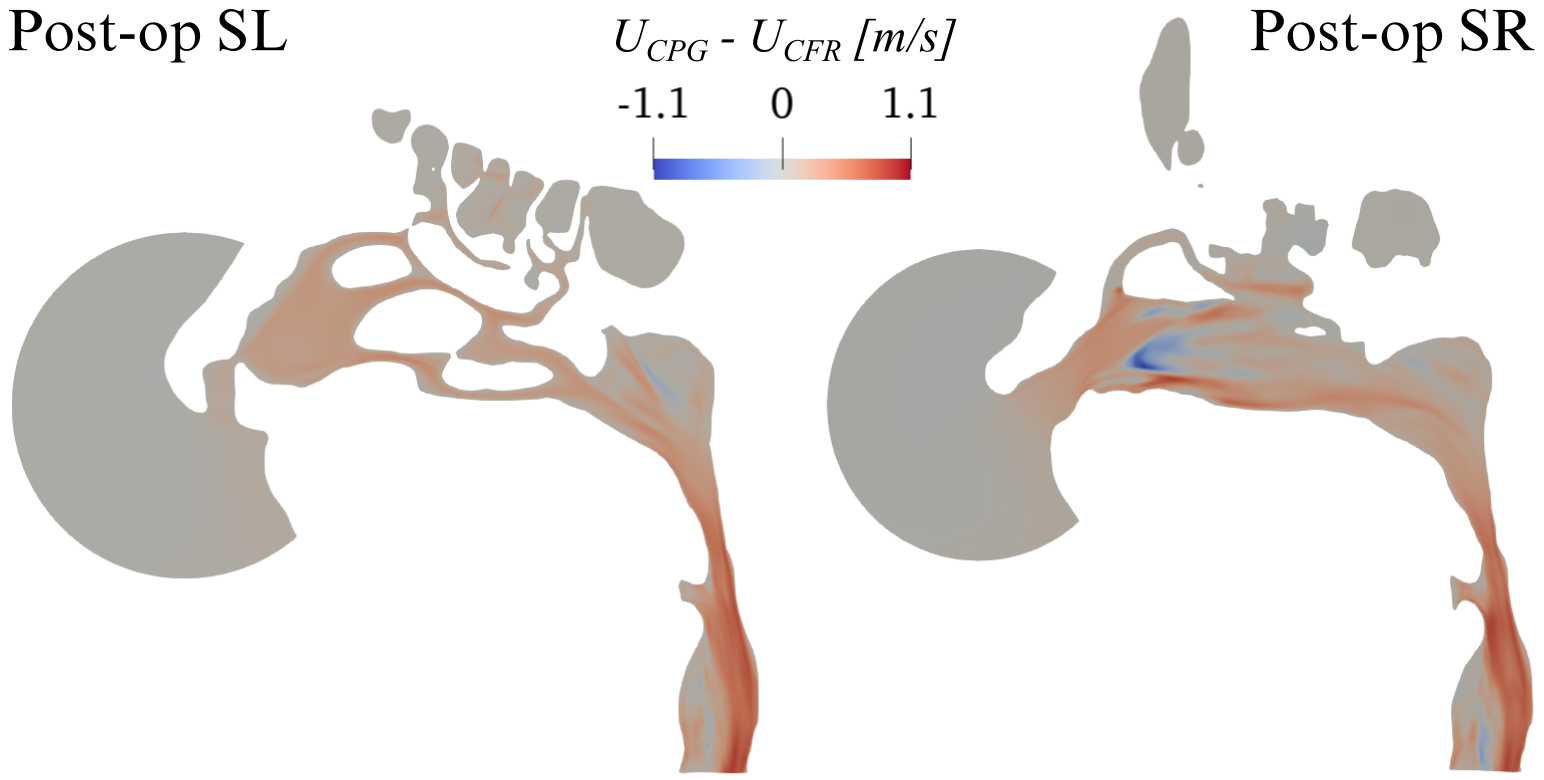}
\caption{Post-op changes of the magnitude $U$ of mean velocity between CPG and CFR. Plot of $U_{CPG} - U_{CFR}$ in planes SL and SR.}
\label{fig:CPG-CFR-S}
\end{figure}

Figure \ref{fig:CFR-S} compares the magnitude $U$ of the mean velocity for the pre-op and post-op anatomies. 
At a glance, results appear in line with those from CPG. However, while the pre-op cases are, as expected, virtually identical, the post-op ones do show differences. 
These are quantified in Figure~\ref{fig:CPG-CFR-S}, where the local difference between the magnitude of the mean velocity, computed with CPG and CFR, is plotted in planes SL and SR. 
Large local differences are observed, higher than $1 \ m/s$; they are mostly found on the operated SR side, but also the unaltered SL side differs. The latter shows a rather uniform change, descending from a change of flow rate, which extends down to the rhinopharynx; the former, instead, additionally shows important localized differences, with changing sign over a small area. Notably, even though the post-op CPG flow rate is higher than the CFR one, the quantity $U_{CPG} - U_{CFR}$ becomes locally negative. 

\subsection{Constant Power Input (CPI)}
In a CPI simulation, the flow is driven by a time-varying pressure drop, and achieves an equally time-varying flow rate; both quantities are adjusted in time to instantaneously achieve the target value of the mechanical power $P=P_0$.
CPI simulations of the flow in the human nose are reported in this paper for the first time.

In the pre-op case, the numerical value for the power input, given by the product of time-averaged flow rate and enforced pressure drop in the pre-op CPG simulation (or, equivalently, by the product of the enforced flow rate and time-averaged pressure drop in the pre-op CFR simulation) is $6.53 \ mW$. The obtained values for the pressure drop and flow rate match, as they should, those of the CPG and CFR simulations. If the same value of power is enforced for a CPI comparison, the post-op simulation yields a 9.5\% reduction of the pressure drop, accompanied by a 10.5\% increase of the flow rate.

\begin{figure}
\centering
\includegraphics[width=\columnwidth]{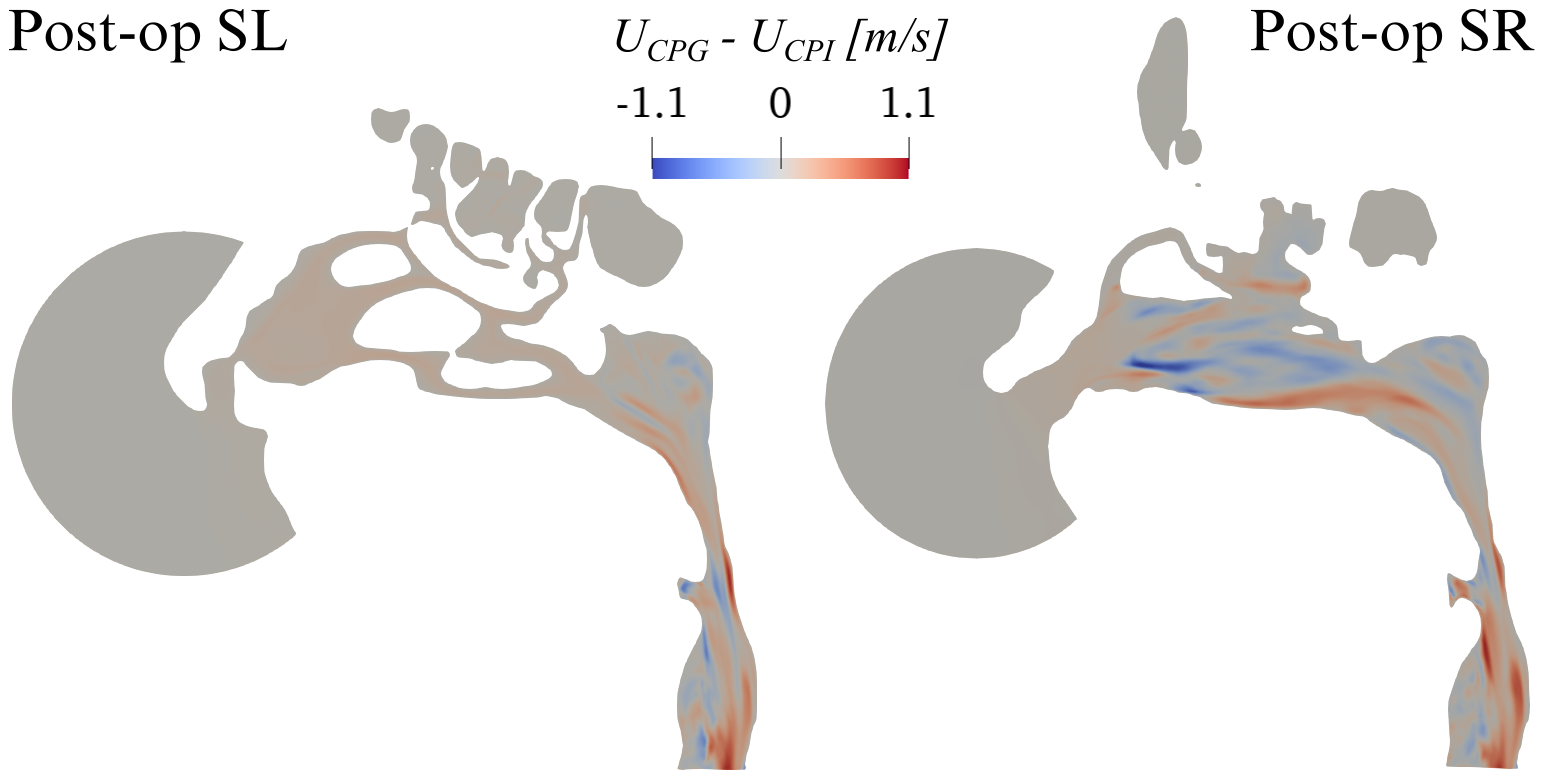}
\caption{Post-op changes in the magnitude $U$ of mean velocity between CPG and CPI. Plot of $U_{CPG} - U_{CPI}$ in planes SL and SR.}
\label{fig:CPG-CPI-S}
\end{figure}

Figure \ref{fig:CPG-CPI-S} compares the magnitude of the post-op mean velocity computed with CPG and CPI, by plotting $U_{CPG} - U_{CPI}$ in planes SL and SR. 
Noticeable differences on both sides are evident, which qualitatively resemble those discussed in figure \ref{fig:CPG-CFR-S} for CFR. 

\section{Discussion}
\label{sec:discussion}

The results presented above demonstrate that post-op velocity and pressure fields depend significantly upon the choice of the flow forcing, in terms of both global and local quantities. 
What forcing to choose remains to some extent a free decision, but one to be taken consciously; the main goal of the present contribution is to highlight the implications of this important logical step.
This issue has gone essentially unnoticed so far, for the main reason that computing a single case with either CFR or CPG is essentially equivalent; differences only appear when a comparison between two cases has to be made. 
The same applies to CPI, a third alternative introduced in this work for the first time in the context of biological flows. 

\begin{figure}
\centering
\includegraphics[width=\columnwidth]{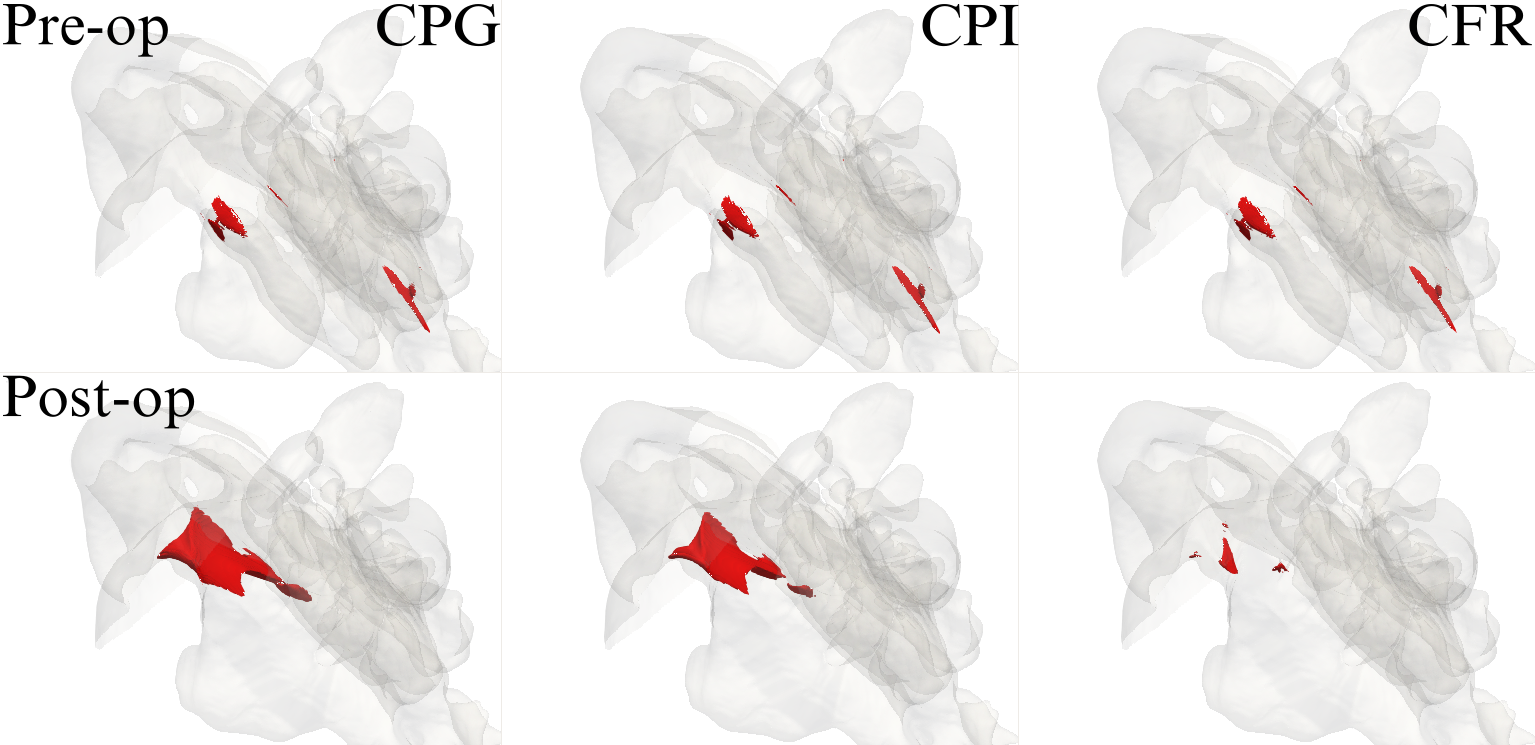}
\caption{Magnitude $U$ of the mean velocity, for all computed cases, in a three-dimensional view. The iso-surfaces correspond to the level of $U = 3 \ m/s$.}
\label{fig:UMean-isoSurfaces-all}
\end{figure} 

Figure~\ref{fig:UMean-isoSurfaces-all} shows isosurfaces for the magnitude of the mean velocity vector, computed with all the forcing strategies, and for the pre- and post-op cases. Obviously, contours are identical in all the pre-op cases (top row), but large and significant differences arise post-op (bottom row). 

\begin{figure}
\centering
\includegraphics[width=0.49\columnwidth]{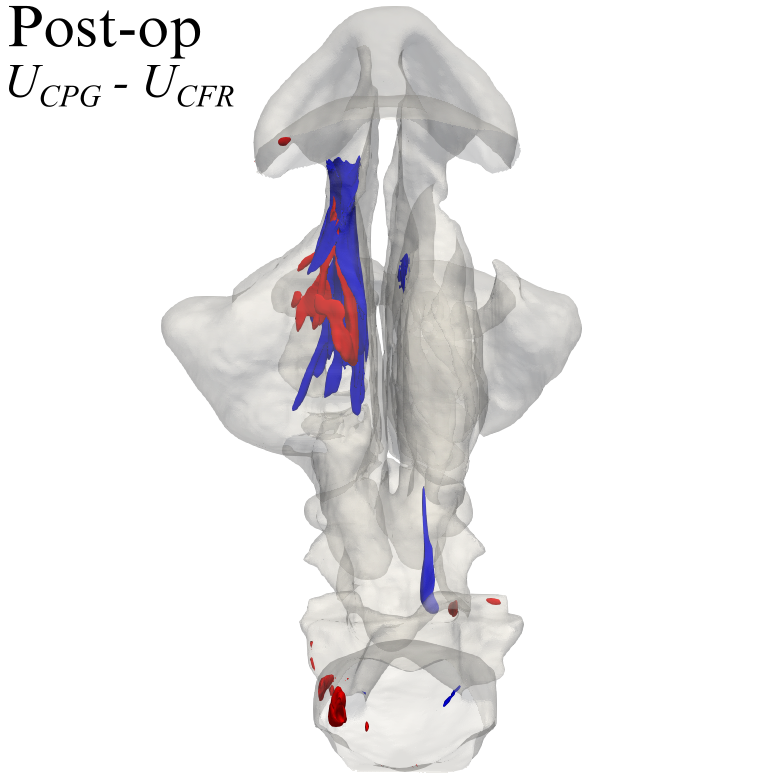}
\includegraphics[width=0.49\columnwidth]{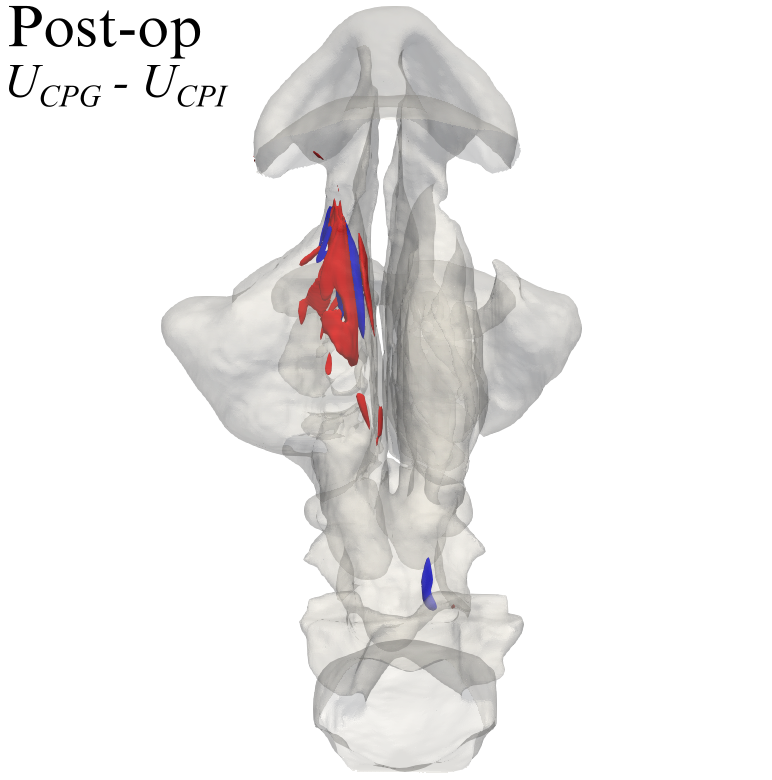}
\caption{Three-dimensional view of the post-op changes in the magnitude $U$ of the mean velocity. Left: $U_{CPG}-U_{CFR}$; right: $U_{CPG}-U_{CPI}$. The red/blue iso-surfaces correspond to $+0.35 \ m/s$ (red) and $-0.22 \ m/s$ (blue) respectively.}
\label{fig:changes-3d}
\end{figure}

Figure \ref{fig:changes-3d} provides a compact three-dimensional view of such differences, in terms of $U_{CPG}- U_{CFR}$ (left) and $U_{CPG}- U_{CPI}$ (right): velocity differences reach up to $\pm 1 \ m/s$, and are particularly significant in the whole right meatus, as a direct consequence of the surgically modified anatomy, and in the rhinopharynx, as an indirect effect of the flow exiting the meatal volumes at different rates. Once again, we notice that differences can take either sign throughout the volume.

We have shown that not only flow details, but global quantities too are affected by the comparison strategy. In fact, flow rate and pressure drop exhibit large relative changes when the flow forcing is changed; power input even changes sign altogether, being increased under CPG and decreased under CFR.
A summary of the numerical values of global quantities measured in the numerical experiments is presented in Table~\ref{tab:resumeResults}.

\begin{table*}
\centering
\caption{Global quantities computed with the three flow forcings.}
\label{tab:resumeResults}
\setlength\tabcolsep{3pt}
\begin{tabular*}{\textwidth}{@{\extracolsep{\fill}}lccccccccc@{\extracolsep{\fill}}}
\toprule
 & \multicolumn{3}{@{}c@{}}{CPG} & \multicolumn{3}{@{}c@{}}{CFR} & \multicolumn{3}{@{}c@{}}{CPI} \\
\cmidrule{2-4}\cmidrule{5-7}\cmidrule{8-10}%
 & pre & post & $\%\Delta$ & pre & post & $\%\Delta$ & pre & post & $\%\Delta$ \\
\midrule
$p_{th} \ [Pa]$  & $-24.4$ & $-24.4$ & - & $-24.4$ & $-18.5$ & $-24.6$ & $-24.4$ & $-22.1$ & $-9.5$ \\
$Q \ [10^{-4} m^3/s]$  & $2.67$ & $3.12$ & $16.9$ & $2.67$ & $2.67$ & - & $2.67$ & $2.95$ & $10.5$ \\
$P \ [10^{-3} W]$  & $-6.53$ & $-7.63$ & $16.9$ & $-6.55$ & $-4.94$ & $-24.6$ & $-6.53$ & $-6.53$ & - \\
$R \ [10^4 Pa \ s/m^3]$  & $9.16$ & $7.84$ & $-14.4$ & $9.18$ & $6.93$ & $-24.5$ & $9.16$ & $7.50$ & $-18.1$ \\
\botrule
\end{tabular*}
\end{table*}

A first important point to remark is that comparing different anatomies is a logical step that becomes relevant not only when evaluating virtual surgeries, but also in the more fundamental search for the functionally normal nose. 
As an example, we mention the study by \cite{zhao-jiang-2014}, in which 22 patients with a normal nasal airflow were compared under the CPG condition. 
When trying to assess which anatomical differences within a number of individuals imply functional consequences, a flow forcing must be selected for the multi-patient comparison, and the outcome is affected by that choice. 
Based on the present results, the conclusions drawn by such studies should be carefully checked to be robust with respect to the type of flow forcing.

The effects observed in the numerical experiments considered here are quantitatively significant, as the endoscopic medial maxillectomy surgery employed as a testbed is a clinically representative surgical maneuver. 
In general, as already observed in turbulent flow control \citep{hasegawa-quadrio-frohnapfel-2014}, CFR and CPG are the extreme cases, with CPI occupying an intermediate position. 
The two most commonly employed forcing strategies, namely CPG and CFR, evidence large and significant differences in global quantities; a 16.9\% increase in flow rate for CPG, and a 24.6\% reduction in pressure drop for CFR. In these two cases, changes in power input have a different sign, with a 24.6\% reduction for CFR and a 16.9\% increase for CPG. 

Table~\ref{tab:resumeResults} also reports values of the nasal resistance $R$, defined as the ratio between pressure drop (between the external ambient and the nasopharynx) and flow rate:
\[
R = \frac{\Delta p}{Q} .
\]

The quantity $R$ is expected to decrease after a surgery like maxillectomy, which enlarges the cross-sectional area of the meati. Indeed, this is found to be the case, regardless of the forcing type. However, if the outcome of the surgery is evaluated through the quantitative change in nasal resistance, the pre-/post-op comparison criterion affects its estimate significantly, with the post-op reduction of $R$ being overestimated by a relative 70\% when computed with CFR (24.5\% reduction) than with CPG (14.4\% reduction). The effect is even larger if one computes the lateral resistances.

Global differences obviously result from the integrated effect of local differences in the flow fields; large velocity differences are found in different parts of the upper airways, as seen in figure \ref{fig:changes-3d}. The major differences are located in the operated right side, but the unoperated airway too shows visible differences. Moreover, the increase in flow rate in the CPG case might lead to changes in position or onset of transition from laminar to turbulent flow, which would affect heat transfer and particle deposition. 
 
\begin{figure}
\centering
\includegraphics[width=\columnwidth]{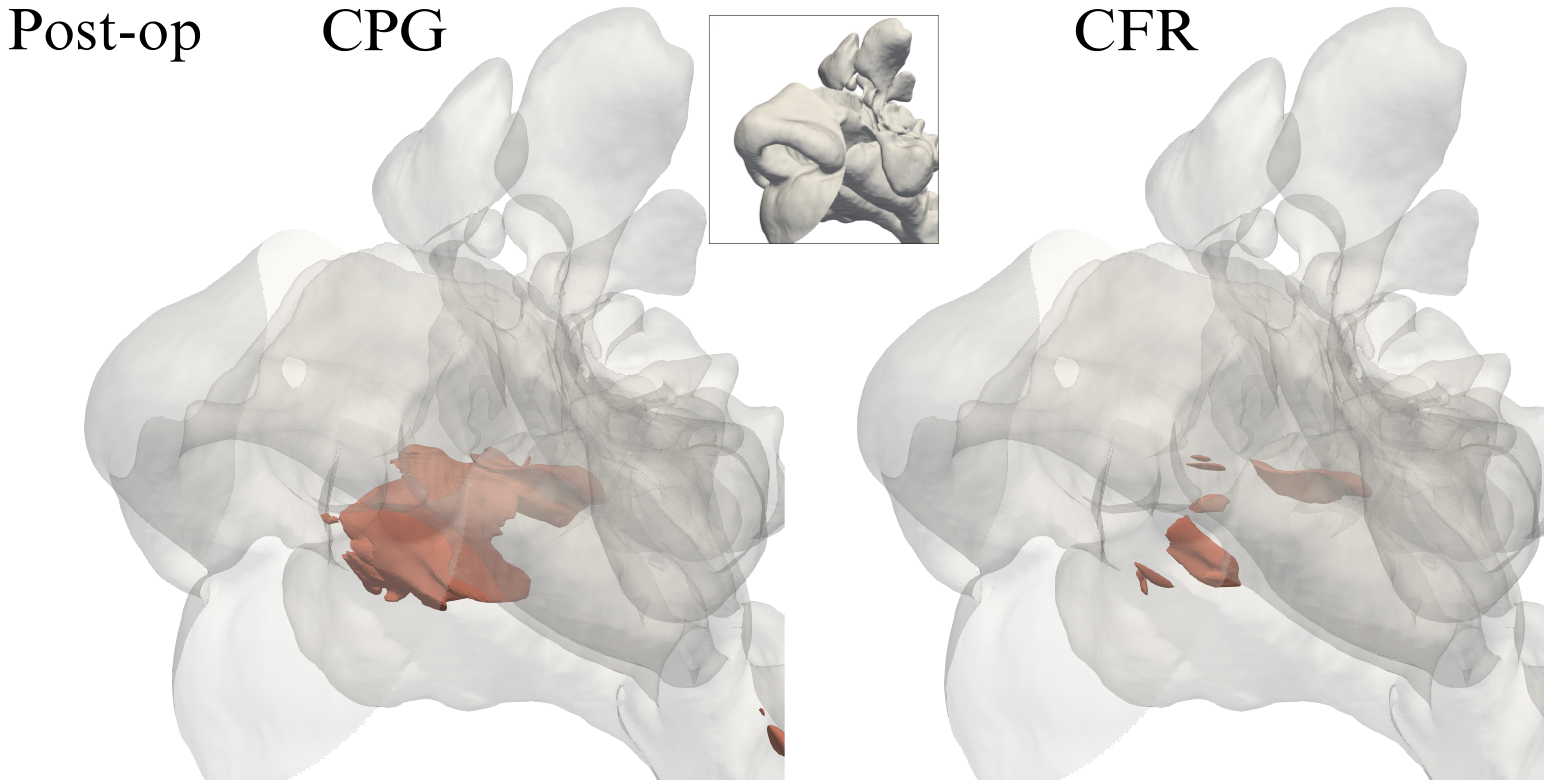}
\caption{Three-dimensional view of the post-op field of turbulent kinetic energy $k$, computed with CPG (left) and CFR (right), and visualized via the iso-surface at the level $k=0.25 \ m^2/s^2$. Zoom in a subregion shown in the inset.} 
\label{fig:tke}
\end{figure} 

Not only the mean field is affected, but turbulent quantities too are found to depend on the flow forcing. As an example, figure \ref{fig:tke} shows the field of the turbulent kinetic energy $k$, and compares the post-op solutions computed with CPG and CFR. One notices a clearly different level of turbulent activity, concentrated right where the (virtual) surgery has modified the anatomy, and significantly larger for the CPG case. It follows that the flow forcing should be considered with extreme care, should CFD be used to assess the outcome of a virtual surgery in terms of the intensity of the turbulent motions.

The last point we put up for discussion is perhaps the most important, and concerns our current inability to fully answer the question set out in the Introduction: How should a comparison be carried out? 
In fact, recognizing the importance and quantifying the effects of the choice of the three forcing strategies is an essential step, but {\em per se} does not lead to the right recipe. In other words, we are still unable to identify the {\em right} forcing, or at least the {\em best} forcing.
Nevertheless, we believe that CPG represents the least adequate choice, on the ground that the value of the enforced pressure drop directly depends on the location of the outlet, hence on the CT scan. Since it is difficult to have precise control on the boundaries of the scan, using CPG would at least require to identify and then stick to one anatomical landmark (e.g. the larynx) and to define the outflow boundary accordingly.
A stronger physiological basis exists to support CFR, which implies comparing breathing at the same rate of oxygen consumption. CPI too, proposed here for the first time, inherits the same difficulties of CPG, but is grounded upon a rather convincing physiological criterion for the comparison, as it implies comparing breathing under the constraint that the same mechanical pressure is provided by the lungs for ventilation. 
The minor disadvantage of CPI, i.e. being not currently available in most commercial CFD solvers, might be more than compensated by its physical appeal, that extends to other field in biomechanics, e.g. when assessing the importance of aneurysms, thromboses or stenoses in blood vessels under the same mechanical power provided by the heart.

\section{Conclusions}
\label{sec:conclusions}

The present work has discussed the implications of choosing the force that drives the flow through the nasal airways when CFD is used to compare two nasal anatomies. Results are obtained for steady inspiration of mild intensity, by using state-of-the-art, well resolved Large Eddy Simulations. 
A pair of pre-op and post-op anatomies has been considered, but the same line of reasoning applies to any two anatomies, e.g. when one is seeking to take advantage of CFD to investigate the functionally normal nose. Similarly, our conclusions apply to any type of comparison, regardless of the specific modelling; although we have employed LES simulations in this work, conclusions apply without modifications to RANS simulations.

A comparison can be carried out at the same pressure drop (CPG), at the same flow rate (CFR), and at the same power input (CPI). 
In particular, the possibility of comparing under the same power input is proposed here for the first time. Beyond the nasal flow, CPI should be considered as a sound alternative in other domains of the fluid dynamics of the human body, e.g. when studying malformations or obstructions of blood vessels (and their surgical corrections), which could be assessed under the same pumping power provided by the heart.
  
The forcing criterion affects the outcome of the comparison in a significant way. For example, variations of nasal resistance induced by surgery change up to a relative 70\%, being largest under CFR and smallest under CPG, with CPI in the middle. Local, instantaneous and time-averaged flow fields are affected as well. 

We have discussed how the CPG approach is, in our opinion, the least adequate choice, owing to the lack of an absolute landmark for the outflow boundary of the computational domain. 
On the other hand, reasonable arguments for both CFR and CPI can be put forward to provide the comparison with a physiological rationale. 
The approaches are equivalent from a computational standpoint, in terms of both complexity and computational cost, although CPI is less straightforward to implement being not immediately available in out-of-the-box commercial solvers.
Choosing between them requires deciding which of the implied physiological constraints is best suited to provide the comparison with clinical significance. Further investigations are needed to arrive at a general community consensus; a clear understanding of the physiological significance of the various boundary conditions might lead to the ability of setting them on a patient-specific basis, e.g. in terms of specific oxygen consumption per unit weight.

\bibliographystyle{sn-chicago}
\bibliography{../../Nose}

\section*{Declarations}

{\bf Funding} The authors did not receive support from any organization for the submitted work.

{\bf Competing interests} The authors have no relevant financial or non-financial interests to disclose.

\end{document}